# Freeform imaging systems: Fermat's principle unlocks 'first time right' design

Fabian Duerr[†] and Hugo Thienpont[‡]

*Brussels Photonics, Department of Applied Physics and Photonics, Vrije Universiteit Brussel*
*Pleinlaan 2, 1050 Brussels, Belgium*

**Abstract:** For more than 150 years, scientists have advanced aberration theory to describe, analyse, and eliminate imperfections that disturb the imaging quality of optical components and systems. Simultaneously, they have developed optical design methods for and manufacturing techniques of imaging systems with ever-increasing complexity and performance up to the point where they are now including optical elements that are unrestricted in their surface shape. These so-called optical freeform elements offer degrees of freedom that can greatly extend the functionalities and further boost the specifications of state-of-the-art imaging systems. However, the drastically increased number of surface coefficients of these freeform surfaces poses severe challenges for the optical design process, such that the deployment of freeform optics remained limited until today.

In this paper, we present a deterministic direct optical design method for freeform imaging systems based on differential equations derived from Fermat's principle and solved using power series. The method allows calculating the optical surface coefficients that ensure minimal image blurring for each individual order of aberrations. We demonstrate the systematic, deterministic, scalable and holistic character of our method with catoptric and catadioptric design examples. As such we introduce a disruptive methodology to design optical imaging systems 'first time right', we eliminate the 'trial and error' approach in present-day optical design, and we pave the way to a fast-track uptake of freeform elements to create the next-generation high-end optics.

## Introduction

Optical imaging systems have been playing an essential role in scientific discovery and societal progress for several centuries[1,2]. For more than 150 years scientists and engineers have used aberration theory to describe and quantify the deviation of light rays from ideal focusing in an imaging system, and to develop methods to design diffraction-limited imaging systems[3]. Until recently most of these imaging systems included spherical and aspherical refractive lenses (dioptric systems) or reflective mirrors (catoptric systems) or a combination of both (catadioptric systems). The last decennia, with the introduction of new ultra-precision manufacturing methods, such as single-point diamond turning and multi-axis polishing[4], two-photon polymerization[5,6] or other additive manufacturing technologies[7], it has become possible to fabricate lenses and mirrors that have at least one optical surface that lacks the common translational or rotational symmetry about a plane or an axis. Such optical components are called freeform optical elements[8] and they can be used to greatly extend the functionalities, improve performance, and reduce volume and weight of optical imaging systems that are principal parts of spectrometers[9], telescopes[10,11], medical imaging systems[12,13], augmented and virtual reality systems[14], or lithography platforms[15,16]. As such imaging systems including freeform optical elements will be key to tremendously advance science and engineering in a wide range of application domains such as astronomy, material research, chip fabrication, visualization, metrology and inspection, energy production, safety and security, biotechnology and medical imaging and diagnosis[17].

Today the design of optical systems largely relies on efficient raytracing and optimization algorithms for which a variety of commercial software[18] and optimization algorithms[19] are available. During an optical design cycle, different parameters of the optical system such as the optical material parameters, radii, coefficients and positions of the optical surfaces are varied to optimize a defined merit function that indicates the image quality for a given field of view[20]. These merit functions are typically "wild" with many local minima, and there is no guarantee that local or global optimization algorithms will lead to a satisfactory design solution[21]. A successful and widely used optimization-based optical design strategy therefore consists of choosing a well-known optical system as a starting point (e.g. from literature) and steadily achieving incremental improvements. Such a "step-and-repeat" approach to optical design, however, requires considerable experience, intuition and guesswork, which is why it is sometimes referred to as "art and science"[22]. This applies especially to freeform optical systems. With the potential of freeform optics widely recognized and its advantages essential to the creation of next generation ultra-performing optical systems, it has become a top priority to overcome the laborious and hardly reproducible trial-and-error approaches currently used for their design.

## Freeform optical design strategies

So far four main design strategies for freeform optical systems have been proposed and developed. All of them target to guide researchers and optical designers to a sufficiently good freeform design that, if needed, can then be used as starting point for subsequent design optimization.

A first strategy involves the so-called **direct design methods**. They rely on solving geometrical or differential equations describing the freeform optical system under study to achieve a well-performing initial design[23–31]. Although these methods show clear merits, so far, they lack a straightforward path to increase the number of optical surfaces that can be calculated[25,29,31]. Because of this limitation in scalability, their applicability remained limited until today.

A second strategy focuses on **automated design and advanced optimization methods** to generate high-performance freeform systems with reduced efforts by the optical designer[32–37]. Although this strategy aims to provide practical design tools to achieve systems with better optical performance, they do not offer valuable insights in the optical design process. Therefore, the cause for not reaching a satisfactory solution often remains unclear, with the only option to restart the process repeatedly.

A third strategy consists in calculating an **initial design that is free of certain aberrations** and then rely on optimization techniques to balance all aberrations and yield best overall imaging performance. One can for example start with an initial rotationally symmetric on-axis design, that has been corrected for several aberrations[38–40], then introduce freeform surfaces and iterate from there, while unobscuring the light path in the system by introducing tilts for the optical surfaces. Alternatively, first-order unobstructed, plane-symmetric systems of three or four spherical surfaces can be calculated as starting points[41,42]. So far, with this design strategy, only a limited number of distinct low-order aberration terms have been cancelled or controlled.

A fourth strategy is based on **nodal aberration theory**[43–45] that has been extended to freeform surfaces[46–48]. This approach allows to predict and visualize the contribution of Zernike terms of an optical surface to the aberration fields and provides information to the designer which Zernike terms are required to correct an aberration of the system[49,50]. So far, nodal aberration theory is one of the more systematic optical freeform design strategies. It can guide the experienced designer towards a successful design provided an in-depth understanding of the underlying theory.

In this paper we present a novel hybrid direct design method for optical imaging systems that also allows to systematically expand the range of aberration terms that need be controlled, suppressed, or cancelled. The method is holistic because it can be used to design imaging systems that include catoptric and/or dioptric spherical, aspherical, and/or freeform surfaces and components. It allows to match user-defined conditions such as minimal blurring for each individual order of aberrations by calculating the corresponding coefficients of the optical surfaces. Moreover, the method is scalable, as it allows to add additional optical surfaces in an unrestricted manner. The method also allows the calculation of aberrations of sufficiently high orders to immediately and accurately estimate the overall imaging performance of the system under design. We highlight the substantial added value for the optical design community by demonstrating the systematic, scalable, deterministic, flexible and holistic character of the design approach with distinct high-end examples, both for catoptric and catadioptric systems. Furthermore, we provide the opportunity for hands-on experience with a design application for imaging systems based on three freeform mirrors. Access to the user application is available upon reasonable request.

## Freeform systems, aberrations, and their mathematical description

There are various possibilities to classify aberrations in freeform optical systems[51,52]. For the design method presented in this paper we distinguish three different categories, according to the overall system symmetry present: (1) systems with two orthogonal planes of symmetry, (2) systems with one plane of symmetry, and (3) systems without any symmetry. Here, we will only consider systems with one plane of symmetry as it is the most common category. The two other symmetry categories can be treated similarly.

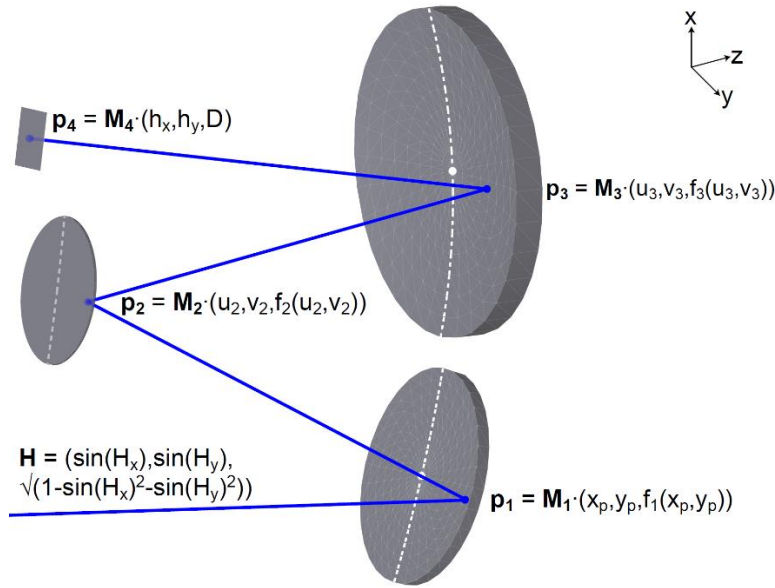

*Fig. 1 Schematic layout of a typical plane-symmetric freeform optical system, here, consisting of three mirrors. The plane of symmetry is indicated by the white dotted lines on each surface. The ray path of an arbitrary ray is shown including the introduced mathematical description of the initial ray direction and intersections with each mirror and the image plane.*

To describe the aberrations in the case of an optical system with one plane of symmetry, for example the x-z plane as illustrated in Fig. 1, we assume an arbitrary ray that is emitted from an object with coordinates $\mathbf{H} = (H_x, H_y, z_0)$, that passes through the pupil of the system (the first optical surface in Fig. 1) at $(x_p, y_p, f1(x_p,y_p))$, and that finally intersects the image plane at $\mathbf{h} = (h_x, h_y, D)$. This intersection can be written in vector form as a series expansion $\mathbf{h} = \mathbf{h_0} + \boldsymbol{\epsilon}^{(1)} + \boldsymbol{\epsilon}^{(2)} + \boldsymbol{\epsilon}^{(3)} + \ldots$ where $\mathbf{h_0} = (g(H_x), g(H_y), D)$ denotes the ideal image location as a function of $H_x$ and $H_y$. Note that $H_x$, $H_y$ can describe a finite as well as an infinitely distant object where $\mathbf{H} = (\sin(H_x), \sin(H_y), (1-\sin(H_x)^2-\sin(H_y)^2)^{1/2})$ is the direction vector corresponding to a field angle of rays entering the optical system. The deviation from the ideal image location is then described by the ray aberration polynomials $\boldsymbol{\epsilon}^{(i)} = (\epsilon_x, \epsilon_y, 0)$

$$\epsilon_x(x_p, y_p, H_x, H_y) = \sum_{j=0}^{\infty} \sum_{k=0}^{\infty} \sum_{l=0}^{\infty} \sum_{m=0}^{\infty} \epsilon_{x,j,k,l,m} x_p^j y_p^k H_x^l H_y^m \quad (1)$$

$$\epsilon_y(x_p, y_p, H_x, H_y) = \sum_{j=0}^{\infty} \sum_{k=0}^{\infty} \sum_{l=0}^{\infty} \sum_{m=0}^{\infty} \epsilon_{y,j,k,l,m} x_p^j y_p^k H_x^l H_y^m \quad (2)$$

in x- and y-direction, respectively, where $(j + k + l + m)$ denotes the aberration order. It is important to remark that the number of independent and non-vanishing aberration coefficients per order depends on the considered symmetry of the freeform optical system[51].

We now consider a sequence of N refractive and/or reflective optical surfaces $f_i$ (i=1…N) (in Fig.1 these are three mirrors) aligned along a principal ray path. To describe their surface functions, we use the power series representation:

$$f_i(x, y) = \sum_{s=0}^{S_i} \sum_{t=0}^{T_i} \frac{1}{s!t!} f_{i,st} x^s y^t \quad (3)$$

In case of one plane of symmetry (here x-z-plane), all coefficients in y with odd *t* vanish. Next we introduce the ray mapping functions $(u_i(x_p, y_p, H_x, H_y), v_i(x_p, y_p, H_x, H_y))$ in x- and y-direction that describe where an arbitrary ray, described by variables $(x_p, y_p, H_x, H_y)$, will intersect each optical surface $f_i$. The ray mapping functions $(u_i, v_i)$ for the $i^{th}$ surface are power series expansions in $(x_p, y_p, H_x, H_y)$ that we write in a similar way as the ray aberration expansions in Eqs. (1) and (2):

$$u_i(x_p, y_p, H_x, H_y) = \sum_{j=0}^{\infty} \sum_{k=0}^{\infty} \sum_{l=0}^{\infty} \sum_{m=0}^{\infty} u_{i,jklm} x_p^j y_p^k H_x^l H_y^m \quad (4)$$

$$v_i(x_p, y_p, H_x, H_y) = \sum_{j=0}^{\infty} \sum_{k=0}^{\infty} \sum_{l=0}^{\infty} \sum_{m=0}^{\infty} v_{i,jklm} x_p^j y_p^k H_x^l H_y^m \quad (5)$$

The chosen principal ray path then determines all vertices of the optical surfaces in a 3D geometry through the series coefficients $(u_{i,0000}, v_{i,0000}, f_{i,00})$. As the surfaces may be tilted and thus not share one common optical axis, the principal ray path direction can change from surface to surface. We can define the respective orientation (tilts) of all elements, namely the object, the individual surfaces of the optical elements, and the image plane, through appropriate rotation matrices $\mathbf{M_H}$, $\mathbf{M_i}$ and $\mathbf{M_h}$, applied to $\mathbf{H}$, $\mathbf{p_i}=(u_i, v_i, f_i)$ and $\mathbf{h}$, respectively. Consequently, the linear surface coefficients $f_{i,10}$ and $f_{i,01}$ are zero for all surfaces. The ray path from the object to the image plane then consists of (N+1) segments $d_1 \ldots d_{N+1}$ that describe the intermediate optical path length distances for an arbitrary ray. Vector geometry and the Pythagorean theorem enable us to express these distances weighted by the refractive indices $n_{i,i+1}$ of the surrounding materials. For example, the distance $d_i$ is calculated as $n_{i-1,i} \| \mathbf{M_i \cdot p_i} - \mathbf{M_i \cdot p_{i-1}} \|$ where $n_{i-1,i}$ denotes the refractive index of the material between two consecutive surfaces and $\|\ldots\|$ is the Euclidean norm.

Let us consider a fixed arbitrary point on the object (finite or infinite object distance), and a fixed but arbitrary point on the second surface $\mathbf{p_2} = \mathbf{M_2} \cdot (u_2, v_2, f_2(u_2, v_2))$. A ray emerging from the object and passing through $\mathbf{p_1}$ towards $\mathbf{p_2}$ must be such that the total optical path length $d_1 + d_2$ is an extremum according to the modern formulation of Fermat's principle.

With points at the boundaries kept fixed, the only remaining variables that can be changed to reach an extremum are $u_1$ and $v_1$ at the point $\mathbf{p_1}=\mathbf{M_1}\cdot(u_1,v_1,f_1(u_1,v_1))$ on the first (in-between) surface $\mathbf{M_1}$. Fermat's principle thus implies that $D_{1x} = \partial_{u_1}(d_1+d_2) = 0$ and $D_{1y} = \partial_{v_1}(d_1+d_2) = 0$. Following similar arguments, we can derive two sets of differential equations for all defined distances $d_i$ pairwise from object to image space

$$D_{i,x} = \partial_{u_i}(d_i+d_{i+1})=0 \quad (i=1\ldots N) \tag{6}$$

$$D_{i,y} = \partial_{v_i}(d_i+d_{i+1})=0 \quad (i=1\ldots N) \tag{7}$$

An optical system consisting of N surfaces is thus fully described by N differential equations $D_{i,x}$ and N differential equations $D_{i,y}$ for i=1...N, for a given but arbitrary pupil plane.

Suppose that all functions $h_x$, $h_y$, $u_i$, $v_i$ and $f_i$ are analytic and smooth solutions of the differential equations $D_{i,x}$ and $D_{i,y}$, then Taylor's theorem implies that the functions must be infinitely differentiable and have power series representations as defined in Eqs. (1) - (5). We can employ a power series method[53] to find solutions to the derived differential equations $D_{i,x}$ and $D_{i,y}$. This method substitutes the power series representations of Eqs. (1) - (5) into the differential equations in Eqs. (6) and (7) in order to determine the values of the series coefficients. To calculate certain coefficients, we differentiate Eqs. (6) and (7) with respect to $x_p, y_p, H_x, H_y$ and evaluate at $x_p = y_p = H_x = H_y = 0$

$$\lim_{x_p \to 0} \lim_{y_p \to 0} \lim_{H_x \to 0} \lim_{H_y \to 0} \frac{\partial^j}{\partial x_p^j}\frac{\partial^k}{\partial y_p^k}\frac{\partial^l}{\partial H_x^l}\frac{\partial^m}{\partial H_y^m} D_{i,x} = 0 \quad (i = 1 \ldots N) \tag{8}$$

$$\lim_{x_p \to 0} \lim_{y_p \to 0} \lim_{H_x \to 0} \lim_{H_y \to 0} \frac{\partial^j}{\partial x_p^j}\frac{\partial^k}{\partial y_p^k}\frac{\partial^l}{\partial H_x^l}\frac{\partial^m}{\partial H_y^m} D_{i,y} = 0 \quad (i = 1 \ldots N) \tag{9}$$

The initial value problem for j=k=l=m=0 is immediately fulfilled and corresponds to the principal ray path that has been defined through all elements' positions and respective rotation matrices. It is important to remark that Eqs. (8) and (9) correspond to the x-components and the y-components of the mapping functions and ray aberrations in the local image plane.

## Deterministic freeform optical design and system evaluation

A preparatory step in the design of a freeform optical system is to specify the layout, number and types of surfaces to be designed and the location of the stop. If we define for example the i$^{th}$ surface as the stop, then $(u_i, v_i, f_i(u_i, v_i))$ will be replaced by $(x_p, y_p, f_i(x_p, y_p))$ in all Eqs. (3) - (9). The layout of the optical system is defined by the path of the chief ray for the central field. The tilts of optical surfaces are entered using rotation matrices. We can now derive the differential Eqs. (6) and (7) and evaluate Eqs. (8) and (9) for the zero-order case for j=k=l=m=0. This condition is mathematically equivalent to the defined chief ray path and is as such automatically fulfilled. At this stage we can include optical specifications such as the entrance pupil diameter (ENPD), focal length (FL), field of view (FOV) and design wavelength as required by the targeted application.

With the differential equations established and the overall system specifications introduced, two design steps need to be taken: (1) solve the non-linear first order case using a standard non-linear solver or by making use of alternative first order optics tools[41,42]; (2) solve the linear systems of equations in ascending order by setting unwanted aberrations to zero or by minimizing a combination thereof as required by the targeted specifications of the imaging freeform system.

These two principal design method steps are identical for all freeform optical designs and are implemented as follows:

(1) We evaluate Eqs. (8) and (9) for all (i,j,k,l,m) for the first order case with j+k+l+m=1. This results in a nonlinear system of equations for the second order surface coefficients $f_{i,st}$ with s+t=2, the mapping coefficients $u_{i,1,0,0,0}$, $u_{i,0,0,1,0}$, $v_{i,0,1,0,0}$, $v_{i,0,0,0,1}$ and the aberration coefficients $\epsilon_{x,1,0,0,0}$, $\epsilon_{x,0,0,1,0}$, $\epsilon_{y,0,1,0,0}$, $\epsilon_{y,0,0,0,1}$. The latter are four first order ray aberration coefficients that can be set to zero or that can be minimized to calculate the unknown surface and the mapping coefficients. Extra conditions for the second order surface coefficients can be imposed if desired. The equations can now be solved using a non-linear solver.

(2) For each of the higher orders o=j+k+l+m=2,3…, we can determine the exact relationship between the surface, mapping and aberration coefficients by evaluating Eqs. (8) and (9) for all corresponding (i,j,k,l,m). This becomes a linear process once the previous order has been solved. Each linear system of order o then relates the surface coefficients $f_{i,st}$ with s+t=o+1 to the mapping coefficients $u_{i,jklm}$, $v_{i,jklm}$ and the aberration coefficients $\epsilon_{x,jklm}$, $\epsilon_{y,jklm}$ with j+k+l+m=o. For each order, we apply the elimination method for solving linear systems to eliminate the unknown mapping coefficients and to obtain a reduced linear system that expresses the aberration coefficients as linear functions of the unknown surface coefficients of that order. If we would calculate the least-squares solution for this reduced linear system, the aberrations of the considered order would be weighted equally. This, however, would not take the system specifications into account. We therefore define a basic set of weighting factors $WF_{j,k,l,m} = (ENPD/2)^{j+k}(FOV/\sqrt{2})^{l+m}$ that is multiplied with the reduced linear equations of same index (j,k,l,m). The weighted least-squares solution[54] for the reduced linear system then takes both the maximum entrance pupil diameter and largest diagonal field diameter into account to simultaneously minimize all properly weighted aberrations for each order, and to calculate the corresponding surface and aberration coefficients. The calculated coefficients are now substituted into the original linear system to calculate the remaining unknown mapping coefficients of that order.

The set of surface coefficients $f_{i,st}$ that have as such been obtained in a deterministic way, now fully describe the N reflective and/or refractive optical surfaces. As such they form a satisfactory 'first time right' solution of the freeform system-under-design, while the aberrations and mapping coefficients can be used to evaluate the imaging quality and the geometry of the optical system. Alternative solutions can be further explored, for example by introducing different weighting factors to rebalance the corrected aberrations, e.g. in favour of a wider field of view.

In addition, the proposed power series method enables a function-based analytic ray tracing evaluation by accurately calculating higher-order mapping and aberration coefficients in ascending order and for all combinations of $j + k + l + m = 1,2,3$ ... up to a desired order that is typically higher than that of the preceding surface coefficient calculations. Thus, the full physical behaviour of the optical system can be immediately interpreted by the optical designer. The obtained 'first time right' solution can be further fine-tuned when introduced in a classical ray-tracing software and applying the embedded advanced optimization algorithms. Fig. 2 summarizes the overall design process in a flowchart, which highlights the difference between input parameters for the optical designer (upper row in blue) and the calculated series coefficients (lower row in orange).

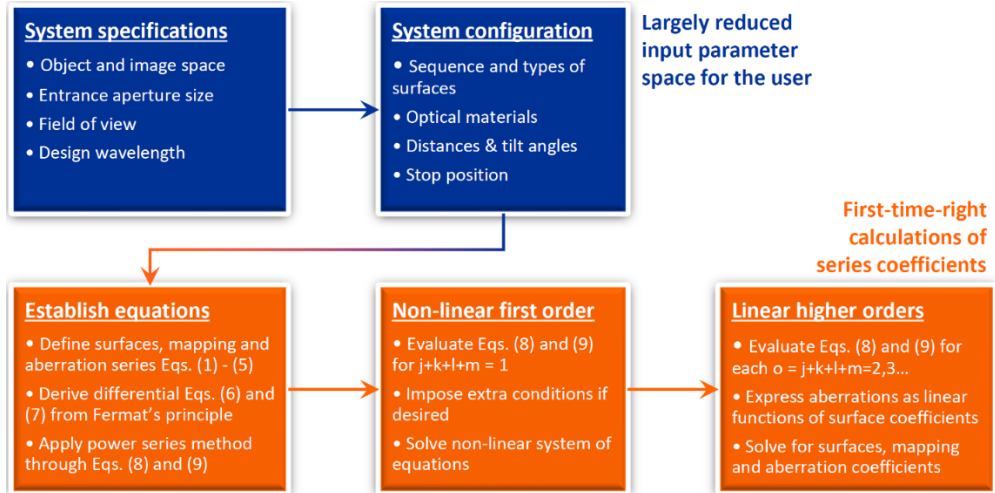

*Fig. 2 Flowchart of the design process, which highlights the required input by the user (upper row in blue) and the directly calculated coefficients by the design method (lower row in orange)*

## Catoptric and catadioptric freeform design: examples and discussion

To illustrate the potential of our deterministic direct optical design method, we apply it to two highly advanced state-of-the-art optical freeform systems that were recently developed by expert groups through elaborate design procedures. To evaluate the design method in real-time we have programmed a graphical user interface (GUI) enhanced MATLAB App. In the shown screenshot in Fig. 3, a partial obscuration is caused by mirror 2, which is directly reflected in a penalty term that is added to the figure of merit function, which should be minimized.

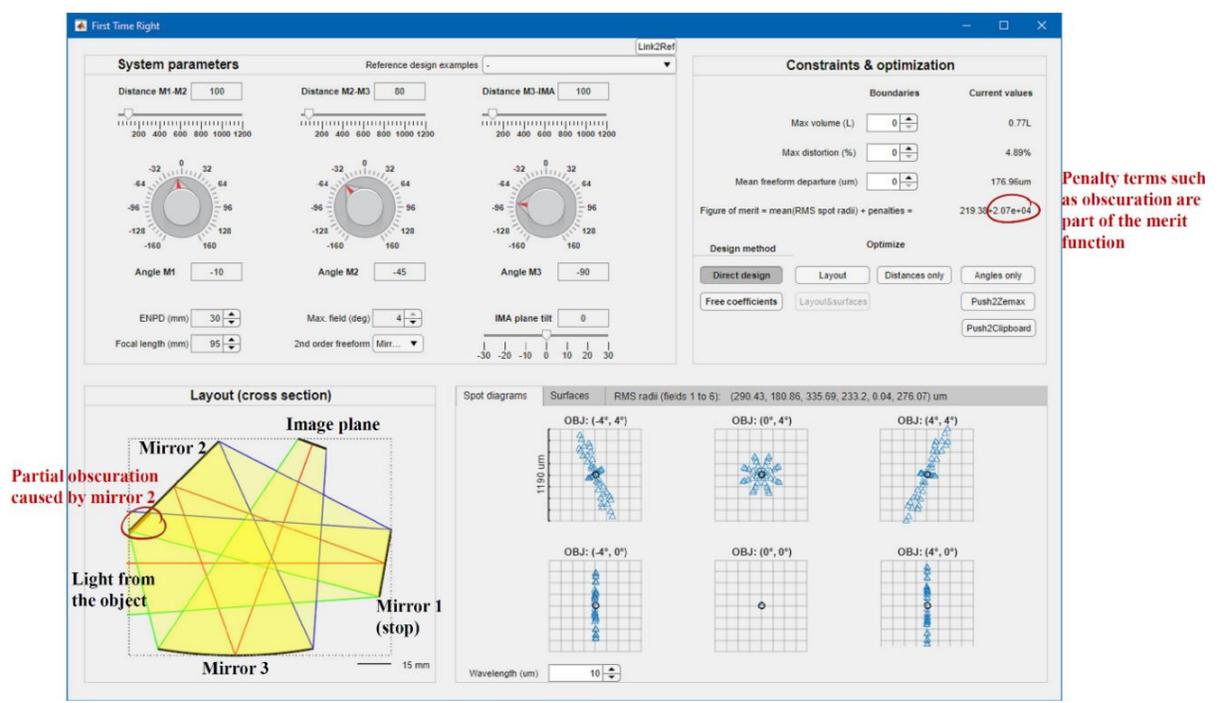

*Fig. 3 Graphical user interface of our developed MATLAB App with overlaid information in the Layout (cross section) window to illustrate the mirror sequence. The partial obscuration caused by mirror 2 (blocking parts of the light from the object) is highlighted and adds a penalty term to the figure of merit.*

The first reference design example is a catoptric imaging system[49] for the visible spectrum with x-z planar symmetry and the object at infinity. It consists of three freeform mirrors where the stop is placed at the first mirror. The targeted system volume is 60 litres. The layout of the optical components can be described by three distances (between mirror 1 and 2, mirror 2 and 3, and mirror 3 and the image plane) and four rotation angles (one for each of the three mirrors, and one for the image plane), which provides seven degrees of freedom for the optical designer. These distances and angles can now be adjusted to create an unobscured starting geometry that meets certain geometrical constraints such as the given target system volume. This can be done manually and in real time within the user application. With the entrance pupil diameter, focal length, field of view and design wavelength specified by the user, we can initiate the mathematical approach as described in the previous section:

(1) Evaluating Eqs. (8) and (9) for all (j+k+l+m) = 1 and i = 1...3 results in a nonlinear system of 12 non-vanishing equations with 18 unknowns. Setting the four first order ray aberration coefficients $\epsilon_{x,1,0,0,0}$, $\epsilon_{x,0,0,1,0}$, $\epsilon_{y,0,1,0,0}$, $\epsilon_{y,0,0,0,1}$ to zero leaves eight mapping coefficients $u_{i,1,0,0,0}$, $u_{i,0,0,1,0}$, $v_{i,0,1,0,0}$, $v_{i,0,0,0,1}$ (i=2,3) and six second order surface coefficients $f_{i,st}$ (s+t=2) as unknowns. We can define two of the three mirrors to have a base curvature, for example $f_{1,20} = f_{1,02} = c_1$ and $f_{3,20} = f_{3,02} = c_3$, reducing the number of unknown surface coefficients from six to four. The nonlinear system can now be solved.

(2) We calculate the surface coefficients up to the sixth order and the mapping and aberration coefficients up to the fifth order, as outlined in the previous section, and continue the linear solution scheme by further increasing the order of the mapping and aberration coefficients to the eighth order (or higher if needed) to ensure a sufficiently accurate prediction of the performance deviation of the system. These calculations result in 40 surface coefficients, 988 mapping and 494 aberration coefficients. We remark here that several of the mapping and aberration coefficients are interdependent.

The calculated design that we obtained by optimizing the seven degrees of freedom of the geometry (three distances and four angles) is shown in Fig. 4a, where the system layout cross-section is combined with the full 3D peak-to-valley freeform departures (PV) from the best-fit base sphere for each mirror, respectively. Fig. 4b shows the resulting spot diagrams for six selected fields based on our aberration calculations. With an average RMS spot radius of about 5 micrometres, our directly calculated system provides an already well-corrected 'first time right' solution that can be readily further optimized.

Next, all forty previously calculated surface coefficients and the initial seven degrees of freedom are used as variables for further optimization, e.g. using MATLAB's *lsqnonlin* solver.

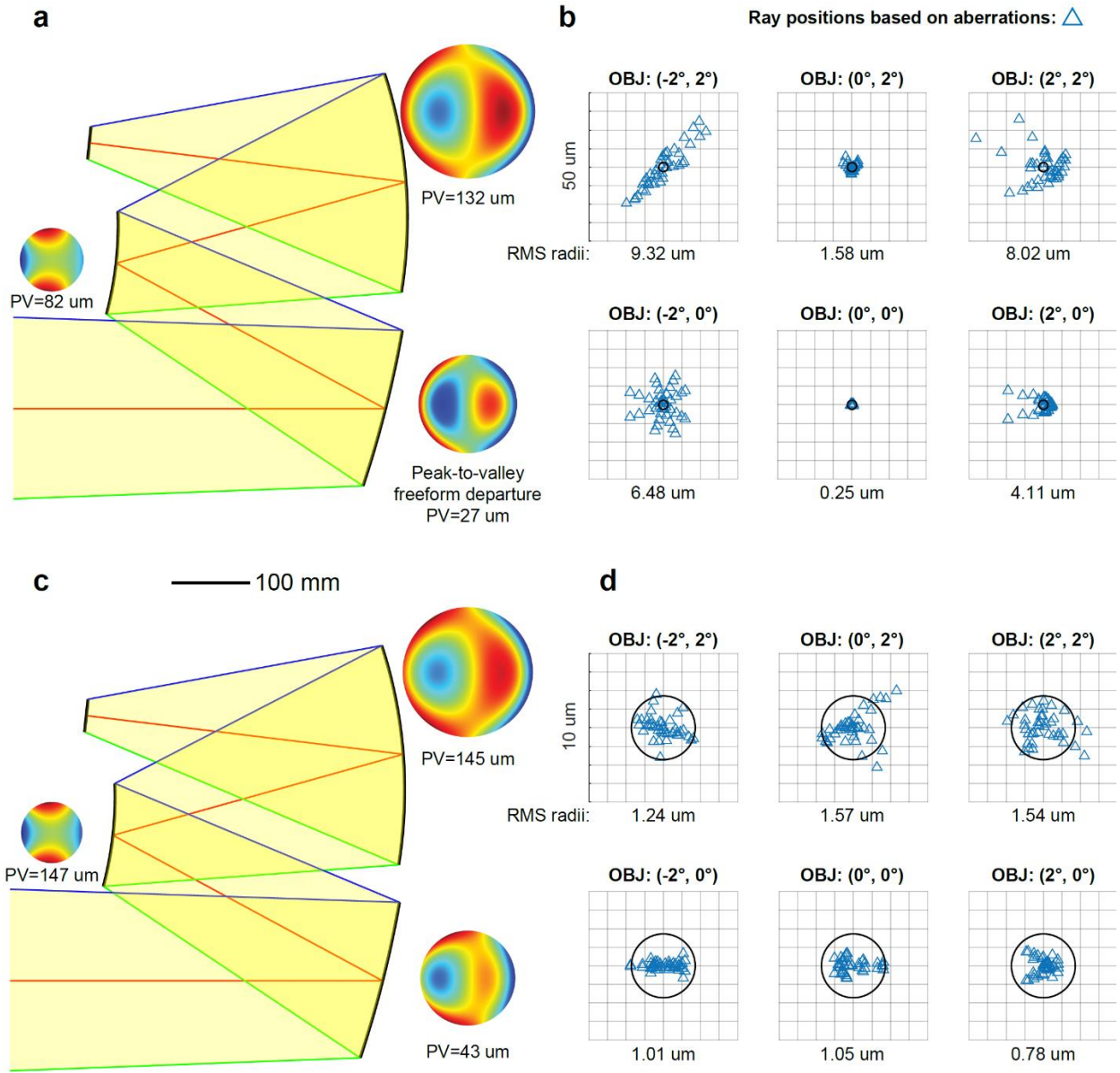

*Fig. 4 'First time right' design results for the three-mirror imager: **a** Cross-section of the directly calculated initial system combined with peak-to-valley freeform departures (PV) from the base sphere for the primary, secondary and tertiary mirror. **b** Corresponding spot diagrams for six selected fields based on aberration calculations. **c** Cross-section of the subsequently optimized system combined with peak-to-valley freeform departures (PV) from the base sphere for each mirror. **d** Corresponding spot diagrams for the same six fields based on aberration calculations.*

After ten iterations, which only take a few minutes of calculation, the system already reaches diffraction-limited performance close to the reported performance[49] for almost the full field of view. The optimized design shows slightly increased and moderate freeform departures distributed among the three mirrors. The results are shown in Fig. 4c and 4d accordingly. The aberration-based performance estimation of our method was found to be in excellent agreement with spot diagram data from classical ray tracing.

It is important to emphasize that these results mainly highlight the very effective nature of this new freeform design and evaluation tool. By using a non-uniform field weighting, and/or a wave front-error-based optimization merit function for the 'first time right' design, a diffraction-limited performance can be readily reached. Most importantly, it becomes evident that the directly calculated system solution provides an excellent starting point that can also be used for subsequent work in a classical design software with advanced optimization algorithms.

As stated in the introduction, increasing the number of calculated surfaces is a major issue for most current direct design methods. As a second example from literature[55], we therefore have selected a monolithic freeform objective for a very compact infrared camera with four optical surfaces, two of which are refractive and two reflective. We chose this example to illustrate the scalability of our design method, well beyond the capabilities of most present-day direct freeform design approaches. We furthermore highlight its holistic character, since in this case we are dealing with a catadioptric system with freeform and aspherical surfaces. Here we design an objective consisting of three aspheres and one freeform at the second surface, with the stop placed at the first surface. It was originally designed by the company Jenoptik AG Jena and discussed by Kiontke[55,56]. An all-freeform redesign of the monolithic objective has been proposed and discussed by Reshidko[57]. We follow these references as closely as possible with the following system requirements: An F/1.4 design covering a 37×25-degree field-of-view, an 8.4 mm entrance aperture, made of optical-grade germanium and operating in the long-wave infrared region (LWIR) from 8-12 um. With the object at infinity, the layout of the system can be described by four distances and five angles, defining the principal ray path from object to image and the respective positions and orientation of all elements and corresponding rotation matrices. It is important to emphasize that the following calculation steps are almost identical to the previous three mirror example, whereas only the non-linear first design step is slightly altered:

(1) We can evaluate Eqs. (8) and (9) for all (j+k+l+m)=1 and for i=1…4, which results in a nonlinear system of 16 non-vanishing equations. We set the four first order ray aberration coefficients $\epsilon_{x,1,0,0,0}$, $\epsilon_{x,0,0,1,0}$, $\epsilon_{y,0,1,0,0}$, $\epsilon_{y,0,0,0,1}$ to zero, leaving 12 mapping coefficients $u_{i,1,0,0,0}$, $u_{i,0,0,1,0}$, $v_{i,0,1,0,0}$, $v_{i,0,0,0,1}$ (i=2,3,4) and eight second order surfaces coefficients $f_{i,st}$ (s+t=2) as unknowns. Here, three surfaces (i=1,3,4) have a base curvature, that is $f_{1,20} = f_{1,02} = c_1$, $f_{3,20} = f_{3,02} = c_3$ and $f_{4,20} = f_{4,02} = c_4$, reducing the number of unknown surface coefficients from eight to five. For the second surface with i=2, a freeform surface, we define the mean base curvature as $c_3 = (f_{3,20} + f_{3,02})/2$, which allows us to apply Petzval's curvature condition[52] for the four curvatures $c_1$ to $c_4$ as a 17th equation. We now solve the nonlinear system.

(2) We calculate the surface coefficients up to sixth order for the surfaces and up to eighth order for the mapping and aberration coefficients.

In order to use a volume close to the reference systems, we constrained the on-axis chief ray path length within the material to 58 mm, a value we estimated through manual measurements. Optimizing the initial degrees of freedom of the geometry yields an already well-corrected and unobscured system. Fig. 5a shows the configuration of the system of comparable size to the cited reference designs. The peak-to-valley freeform departure (PV) from the best-fit base sphere for the second freeform surface is added next to it. Fig. 5b shows the corresponding spot diagrams for six selected fields (we used 15 fields for the optimization) based on our aberration calculations.

With an average RMS spot radius of about 51 micrometres, we achieved an excellent starting point for further optimization. All surface coefficients are now set as variables for further optimization using e.g. MATLAB's *lsqnonlin* solver. The results are shown in Fig. 5c and 5d.

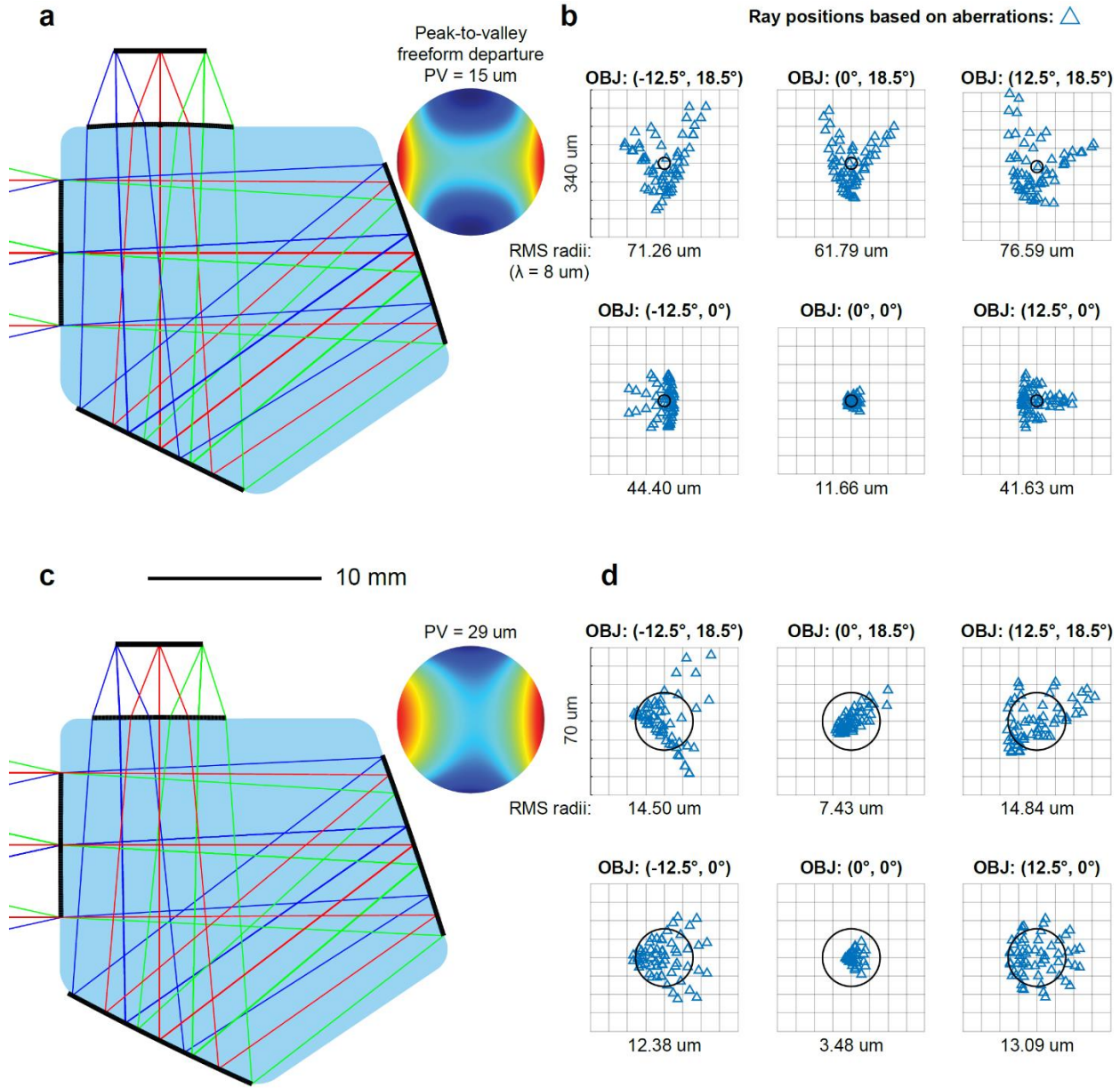

*Fig. 5 'First time right' design results for the four-surfaces monolithic infrared camera objective: **a** Cross-section of the system from direct calculations with peak-to-valley freeform departure (PV) from the base sphere for the second and only freeform surface. **b** Corresponding spot diagrams for six selected fields based on aberration calculations. **c** Cross-section of the subsequently optimized system combined with peak-to-valley freeform departure (PV) from the base sphere for the second surface. **d** Corresponding spot diagrams for the same six fields based on aberration calculations.*

After ten iterations and few minutes, the system reaches a close to diffraction-limited performance in the LWIR band for the full field of view. This spot diagram-based performance is in excellent agreement with spot diagram data from classical ray tracing and seems to exceed the reported performances of the reference systems[55,57]. However, as the authors did not reveal all relevant information such as the nominal focal length, the acceptable distortion (here, less than 5%) or the target volume of their systems, a truly quantitative comparison was not possible.

A comparison with the catoptric three surfaces design example makes clear that the only differences with the catadioptric system designed here are the altered set of underlying Eqs. (6)-(9) and the adapted solution of the non-linear first order equations. Everything else remains the same no matter what types or how many surfaces are used.

This example clearly demonstrates again the highly effective nature of our proposed design and evaluation method. In addition, it highlights two further important features: 1) our method allows to simply combine refractive and/or reflective surfaces of spherical, aspheric or freeform shapes; 2) the method also straightforwardly enables to increase the number of calculated surfaces of an optical system without considerably increasing the computational complexity of the problem. This direct path to scaling the number of calculated surfaces is a result of the fact that the intermediate optical path length distance expressions do not depend on the number of considered surfaces.

## Summary and Conclusions

Equipping lens- and/or mirror-based optical systems with freeform optical surfaces makes it possible to deliver highly original imaging functionalities with superior performance compared to their more traditional (a)spherical counterparts, such as enhanced field-of-view, increased light-collection efficiencies, larger spectral band and higher compactness. Until now mathematical models and design strategies for freeform optics remained limited and failed to provide deterministic solutions. In particular, the identification of a suitable initial design has proven to be a painstaking and time-consuming trial-and-error process.

In this paper, we reported on the first deterministic direct design method for freeform optical systems that is not restricted by the aberration terms that can be controlled. The method relies on Fermat's principle and allows a highly systematic generation and evaluation of directly calculated freeform design solutions that can be readily used as starting point for further and final optimization. As such, this new method allows the straightforward generation of a 'first time right' initial design that enables a rigorous, extensive and real-time evaluation in solution space.

We can summarize the main features of our new method as follows:

(1) **Holistic:** In contrast to current partial solutions based on aberration theory, this method is not restricted to controlling and correcting a limited number of low-order aberrations but works for all present aberrations in freeform optical systems up to the desired order.

(2) **Deterministic:** As many current design strategies strongly rely on multi-step optimization routines, final design results often differ considerably for slightly different starting points. The here presented method allows a fast and deterministic solution as well as a very systematic evaluation of the solution space while providing more detailed insights into the fundamental physics of the optical system than most conventional design approaches.

(3) **Scalable:** So far, several direct design methods have been reported for freeform optical systems. However, these methods have in common, that they are developed for a specific system layout, e.g. two or three freeform mirrors. Once a designer would like to add one or more mirrors to the design, the methods do not typically scale with the number of calculated optical surfaces. In contrast, the here presented method follows a clear set of rules that allows to add additional refractive and/or reflective surfaces to a system.

**(4) All-round:** the method can be beneficial for "junior" and "senior" optical designers alike. The in-depth insights into aberrations with this approach are very valuable for skilled and senior optical designers. Inexperienced optical designers with a less advanced understanding of aberration theory on the other hand can equally benefit from it, as this in-depth insight is not a must-have to make excellent use of the method. Finally, the drastic reduction in parameter space, i.e. the large ratio of calculated coefficients over input coefficients, make this method very attractive for designers with a strong background in optimization.

The deterministic, holistic, and scalable nature of this all-round method therefore has the full potential to create a true paradigm shift in how freeform imaging systems will be designed and developed from now on to suit a wide range of applications.